\documentclass[amsmath,amssymb,reprint]{revtex4-1}

\usepackage{graphicx} 
\usepackage{multirow}
\usepackage{array}
\usepackage{subfigure}

\usepackage{hyperref} 

\begin{document}


\title{Synchronization optimized networks for coupled nearly identical oscillators and their structural analysis}

\author{Suman Acharyya$^1$} 
\email{suman@prl.res.in}
\author{R. E. Amritkar$^{1,2}$} 
\email{amritkar@prl.res.in}
\affiliation{$^1$Physical Research Laboratory, Ahmedabad - 380009, India\\
$^2$Institute of Infrastructure, Technology, Research and Management, Ahmedabad, India}

\begin{abstract}
 The extension of the master stability function (MSF) to analyze stability of generalized synchronization for coupled nearly identical oscillators is discussed. The nearly identical nature of the coupled oscillators comes from some parameter mismatch while the dynamical equations are the same for all the oscillators.  From the stability criteria of the MSF, we construct optimal networks with better synchronization property, i. e. the synchronization is stable for widest possible range of coupling parameter. In the optimized networks the nodes with parameter value at one extreme are selected as hubs. The pair of nodes with larger parameter difference are preferred to create links in the optimized networks. And the optimized networks are found to be disassortative in nature, i. e. the nodes with high degree tend to connect with nodes with low degree.
\end{abstract}


 
\maketitle

\section{Introduction}

Synchronization processes of locally interacting dynamical systems has been the focus of intense research in physical, biological, chemical, technological and social sciences~\cite{sync_ref,PecoraPRL1990,Heagy1994,PecoraPRL1998,Arenas2008}. The simplest and the most studied form of synchronization is the complete synchronization (CS) which is observed when identical dynamical systems are coupled and their state variables become equal as time goes to infinity. As a result of this equality of the state variables the motion of the coupled systems collapses onto a subspace of the overall phase space. This subspace is known as synchronization manifold and the remaining directions in the phase space define the transverse manifold. The complete synchronization is stable when all perturbations in the transverse direction decay with time.

One very important tool for the study of stability of complete synchronization of coupled identical oscillators is the \emph{Master Stability Function} (MSF), introduced by Pecora and Carroll~\cite{PecoraPRL1998}. The MSF is defined as the largest Lyapunove exponent, calculated from a set of equations known as Master Stability Equation (MSE). The MSF simplifies the study of stability for complete  synchronization by separating the effect of the network structure from that of the dynamics of individual systems. The MSF also facilitates the stability analysis of complete  synchronization for coupled identical oscillators by constructing a single function which can be used to compare synchronizability of different networks. 

In practical world, most of the interacting dynamical systems are nonidentical in nature and being nonidentical they cannot exhibit complete synchronization; instead they undergo generalized synchronization~\cite{Abarbanel1995,Abarbanel1996}. Thus, to better understand synchronization processes of natural systems it is important to construct a master stability function for generalized synchronization. Motivated by this problem, we have extended the formalism of the master stability function to the generalized synchronization of coupled nearly identical oscillators~\cite{Acharyya2012,Acharyya2013,Acharyya2014}. 

During the last two decades, the theory of complex network has evolved tremendously~\cite{compnet_ref}. Many real world complex systems can be modelled as complex networks of interacting dynamical oscillators. Recently, the effect of structure of complex network on the synchronization dynamics of coupled oscillators has been investigated~\cite{FernandezPRL2000,Barahona2002,Nishikawa2003,Donetti2005}. In an earlier work, it is shown that the small world scheme enhances synchronizability for a network of coupled identical systems~\cite{Barahona2002}. A general argument underlying this phenomenon is that the communication between the coupled systems is more efficient because of the smaller average network distance. In Ref~\cite{Nishikawa2003} it is shown that having smaller network distance is not sufficient for performing best synchronizability properties, it is also required to have homogeneous degree distribution among the coupled dynamical systems. In Ref.~\cite{Donetti2005}, the authors introduced a new family of graphs, namely the entangled networks, which optimizes synchronizability for many dynamical systems. These entangled networks are interwoven and has an extremely homogeneous structures, i.e. the degree distributions are very narrow. All of the above mentioned results are for networks with identical dynamical oscillators.

The study of finding optimal topology of networks which exhibit better synchronizability has been a subject of paramount interest. Recently, the edge rewiring method has been used vastly for finding optimal topologies of networks for better synchronizability~\cite{Donetti2005,JaliliChaos2008,JaliliChaos2010}. The optimal networks obtained in this way are mostly homogeneous networks, i.e. the  degree distributions of these networks are very narrow. 

We use the stability criteria provided by the master stability function to construct optimal networks which shows better synchronizability for nearly identical systems. In the optimized networks the nodes with parameter value at one extreme are selected as hubs. The pair of nodes with larger parameter difference are preferred to create links in the optimized networks. And the optimized networks are found to be disassortative in nature, i. e. the nodes with high degree tend to connect with nodes with low degree.

\section{Master stability function for nearly identical oscillators}

\begin{figure}
\begin{center}
\includegraphics[]{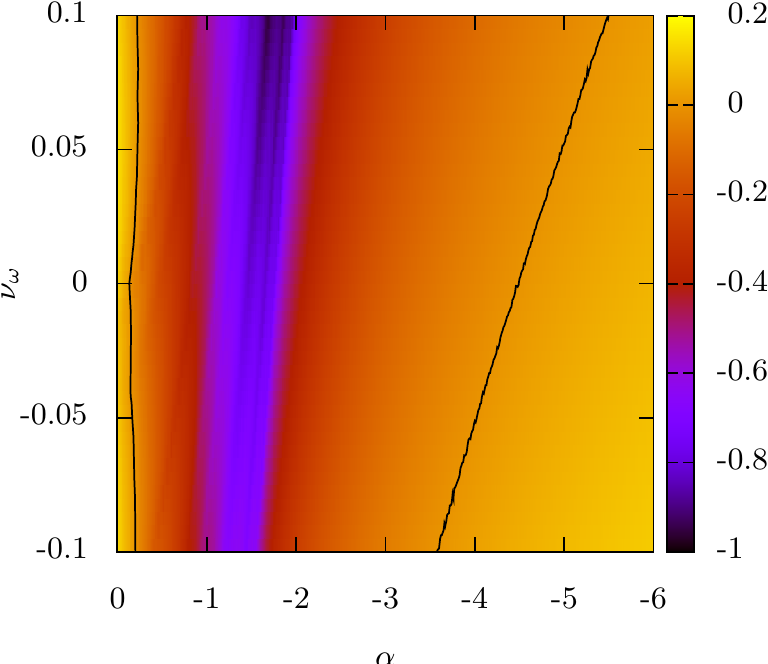}
\end{center}
\caption{\label{msfrosomega_colormap} The colormap of the MSF for coupled nearly identical R\"ossler systems with the frequency $\omega_i$ as the NDP, is shown on the $\alpha$-$\nu_{\omega}$ plane. Here, we take the coupling matrix to be symmetric so that $\alpha$ and $\nu_{\omega}$ are real. The zero values of the MSF are shown by the two curves. We can see that the stability of MSF increases as $\nu_{\omega}$ increases. The other R\"ossler parameters are $a=b=0.2,c=7.0$.}
\end{figure}

We consider a network of $N$ coupled nearly identical chaotic oscillators. The dynamics of $i$-th oscillator is given by
\begin{eqnarray}
\dot{x}^i = f(x^i,r_i) + \varepsilon \sum_{j=1}^N g_{ij} h(x^j),\; i=1,\ldots,N
\label{i_dyn_eq}
\end{eqnarray}
where, $x \in R^m$ is $m$ dimensional state variable and $f: R^m \rightarrow R^m$ provides the dynamics of the isolated oscillator, $\varepsilon$ is scalar coupling strength. $G=[g_{ij}]$ is coupling matrix; if systems $i$ and $j$ interact then $g_{ij}=1, i\neq j$, otherwise $g_{ij}=0$ and the diagonal elements of $G$ are $g_{ii} = -\sum_{j=1;j\neq i}^N g_{ij}$. Thus the elements of $G$ satisfy $\sum_{j}g_{ij} = 0$. $h:R^m \rightarrow R^m$ is a linear coupling function. $r_i$ is some parameter of the dynamics that depends on oscillator $i$. Let $r_i = \tilde{r} + \delta r_i$, where $\tilde{r}$ is a typical value of the parameter $r$ and $\delta r_i$ is a small parameter mismatch. We call the parameter $r_i$ as a \emph{Node Dependent Parameter} (NDP).

When $\delta r_i = 0;\; \forall i$, the coupled oscillators will be identical and for suitable coupling function $h$ and coupling parameter $\varepsilon$ the coupled oscillators will undergo complete synchronization, i.e. $x^i = s(t);\; \forall i$. 

Now, we consider the case where $\delta r_i \neq 0$ and in this case the synchronization between the coupled oscillators will be of generalized type, i.e. there will be a functional relationship between the variables, $\phi(x^1,\ldots,x^N) = 0$.

To determine the stability of generalized synchronization, we do linear stability analysis. Throughout this paper we consider $\delta r_i$ to be small, i.e. $\frac{\delta r_i}{r_i} \ll 1;\; \forall i$. Due to this condition the attractors of the coupled oscillators are not very different from each other. This enable us to expand Eq.~(\ref{i_dyn_eq}) in Taylor series about the solution $\tilde{x}$ of an isolated oscillator with NDP $\tilde{r}$. 

The exponential nature of solution of a linear differential equation is dominated by the homogeneous term of that differential equation. The effect of the NDP appears first in the homogeneous part from the quadratic terms in Taylor series expansion of the function $f(x^i,r_i)$. We retains terms up-to second order in $z^i = x^i - \tilde{x}$ and $\delta r_i = r_i - \tilde{r}$. The dynamics of deviation is given by
\begin{eqnarray}
\dot{z}^i &=& D_x f(\tilde{x},\tilde{r}) z^i + \varepsilon\sum_{j=1}^N g_{ij} D_x h(\tilde{x}) z^j + D_r f(\tilde{x},\tilde{r})\delta r_i  \nonumber \\
& & +  \frac{1}{2}D_x^2 f(\tilde{x},\tilde{r}) (z^i)^2 + D_r D_x f(\tilde{x},\tilde{r}) z^i \delta r_i + \frac{1}{2} D_r^2 f(\tilde{x},\tilde{r}) \delta r_i^2
\label{TE_full}
\end{eqnarray}
As an equation for $z^i$, the RHS of Eq.~(\ref{TE_full}) contains both homogeneous and inhomogeneous terms. To a first approximation, the inhomogeneity of a linear ordinary differential equations does affect the Lyapunov exponents or the exponential rate of convergence or divergence of the solutions though it can shift the solutions \cite{Acharyya2012}. In nonlinear systems, in addition to the shift the attractor may also deform which can lead to a change in the exponent. This is the case in the desynchronized state. However, in the synchronized state attractors of the coupled oscillators are in generalized synchrony and are related to each other, i.e. $\phi(x^1,\ldots,x^j)=0$. Hence it is reasonable to conjecture that in the synchronized state the shifted solution preserves the nature of the attractor so that the average expansion and contraction rates are not significantly affected  \cite{Acharyya2012}.

Hence, to calculate Lyapunov exponent from Eq.~(\ref{TE_full}) we consider the homogeneous equation obtained from Eq.~(\ref{TE_full})
\begin{eqnarray}
\dot{z}^i &=& D_x f(\tilde{x},\tilde{r}) z^i + \varepsilon\sum_{j=1}^N g_{ij} D_x h(\tilde{x}) z^j  + D_r D_x f(\tilde{x},\tilde{r}) z^i \delta r_i 
\label{homogeneous_eq}
\end{eqnarray}
In matrix for Eq.~(\ref{homogeneous_eq}) can be written as
\begin{eqnarray}
\dot{Z} &=& D_x f \ Z + \varepsilon  D_x h \ Z \ G^T + D_r D_x f \ Z \ R. 
\label{homogeneous_eq_mat_form}
\end{eqnarray}
where $G^T$ is the transpose of the coupling matrix $G$ and $R=\text{diag}(\delta r_1,\ldots,\delta r_N)$ is $N\times N$ is a diagonal matrix with the diagonal entries as the mismatch in NDP. We can see from Eq.~(\ref{homogeneous_eq_mat_form}) that it is necessary to include the quadratic terms in $z^i = x^i - s$ and $\delta r_i$ in the Taylor series expansion as the effect of the NDP is not present in the linear terms.

\begin{figure*}
\begin{center}
\includegraphics[]{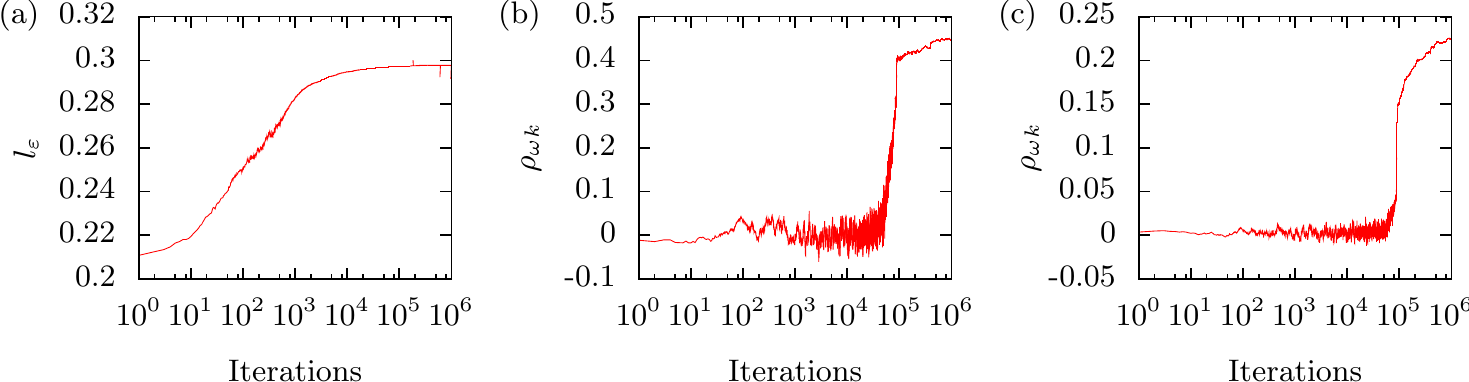}
\end{center}
\caption{\label{leps_rho} (a) The interval $l_{\varepsilon}$ of the coupling parameter for stable synchronization is plotted as a function of Monte Carlo iterations.
(b) The figure plots the correlation coefficient $\rho_{\omega k}$ as a function of the Monte Carlo iterations. We see that $\rho_{\omega k}$ increase and saturate to a positive values.
(c) The figure plots the correlation coefficient $\rho_{\omega A}$ as a function of the Monte Carlo iterations. We see that $\rho_{\omega A}$ increase and saturate to positive values.
In these simulations, we have 32 coupled nearly identical R\"ossler oscillators with the number of edges as $174$. The NDP is $\omega$.
All numerical results are averaged over 100 runs.}
\end{figure*}

Let $\gamma_k$ be the $k$-th eigenvalue and $e_k^R$ be the corresponding right eigenvector of $G^T$. Define an $m$ dimension vector $\eta_k = Z e_k^R$. The dynamics of the vector $\eta_k$ is given by
\begin{eqnarray}
\dot{\eta}_k & = & [D_xf + \varepsilon \gamma_k D_xu] \eta_k + D_r D_x f \ Z \ R \ e_k^R,\; k = 1,\ldots, N.
\label{dyn_eta}
\end{eqnarray}
In general, $e_k^R$ are not eigenvectors of $R$ and hence Eq.~(\ref{dyn_eta}) is not easy to solve. To solve
Eq.~(\ref{dyn_eta}) we use first order perturbation theory~\cite{matrix-computation-book} and write Eq.~(\ref{dyn_eta}) as
\begin{eqnarray}
\dot{\eta}_k & = & [D_xf + \varepsilon \gamma_k D_xu + \nu_k D_r D_x f] \eta_k \label{dyn_eta_pert}
\end{eqnarray}
where $\nu_k = (e_k^L)^T R e_k^R$ is the first order correction and $e_k^R$ and $e_k^L$ are the right and left eigenvectors of $G^T$ corresponding to the eigenvalue $\gamma_k$. 

Since both $\gamma_k$ and $\nu_k$ can be complex, treating them as complex parameters $\alpha = \varepsilon \gamma_k$ and $\nu = \nu_k $ respectively, we can construct the master stability equation as
\begin{equation}
\dot{\eta} = [D_xf + \alpha D_xh + \nu D_rD_x f] \eta. \label{msf_eqn}
\end{equation}
We call $\alpha$ as network parameter and $\nu$ as mismatch parameter.

The master stability function (MSF) is defined as the largest Lyapunov exponent calculated from Eq.~(\ref{msf_eqn}). The stability of the synchronization is given by the negativity of the MSF. 
For coupled identical systems, the above equation reduces to the master stability equation given by Pecora and Carroll \cite{PecoraPRL1998} by setting the mismatch parameter $\nu=0$.

Here we note that the eigenvalue $\gamma_1=0$ of the coupling matrix $G^T$ corresponds to the synchronization manifold and the remaining eigenvalues $\gamma_k;\; k=2,\ldots,N$ correspond to the transverse manifold. For a given network the synchronization is stable when all Lyapunov exponents corresponding to the eigenvalues $\gamma_k;\; k=2,\ldots,N$ of $G^T$ are negative, i.e. they fall in the region where the MSF is negative.

For many chaotic oscillators it is observed that the MSF is negative in a finite interval of the network parameter $\alpha$. Let, the interval be $(\alpha_l,\alpha_h)$ for identical oscillators and $(\alpha_l',\alpha_h')$ for nearly identical oscillators. We can write the condition for stable synchronization of a given network of coupled nearly identical oscillators as
\begin{equation}
 \alpha_l' < \varepsilon\gamma_2\leq \ldots \leq \varepsilon\gamma_N<\alpha_h'
 \label{syn_cond_1}
\end{equation}
The above condition can also be written as
\begin{equation}
\frac{\gamma_N}{\gamma_2} < \frac{\alpha_h'}{\alpha_l'}
\label{syn_cond_2}
\end{equation}

\begin{figure*}[t]
\subfigure[Initial Network]{
\centering
\includegraphics[width=.45\textwidth]{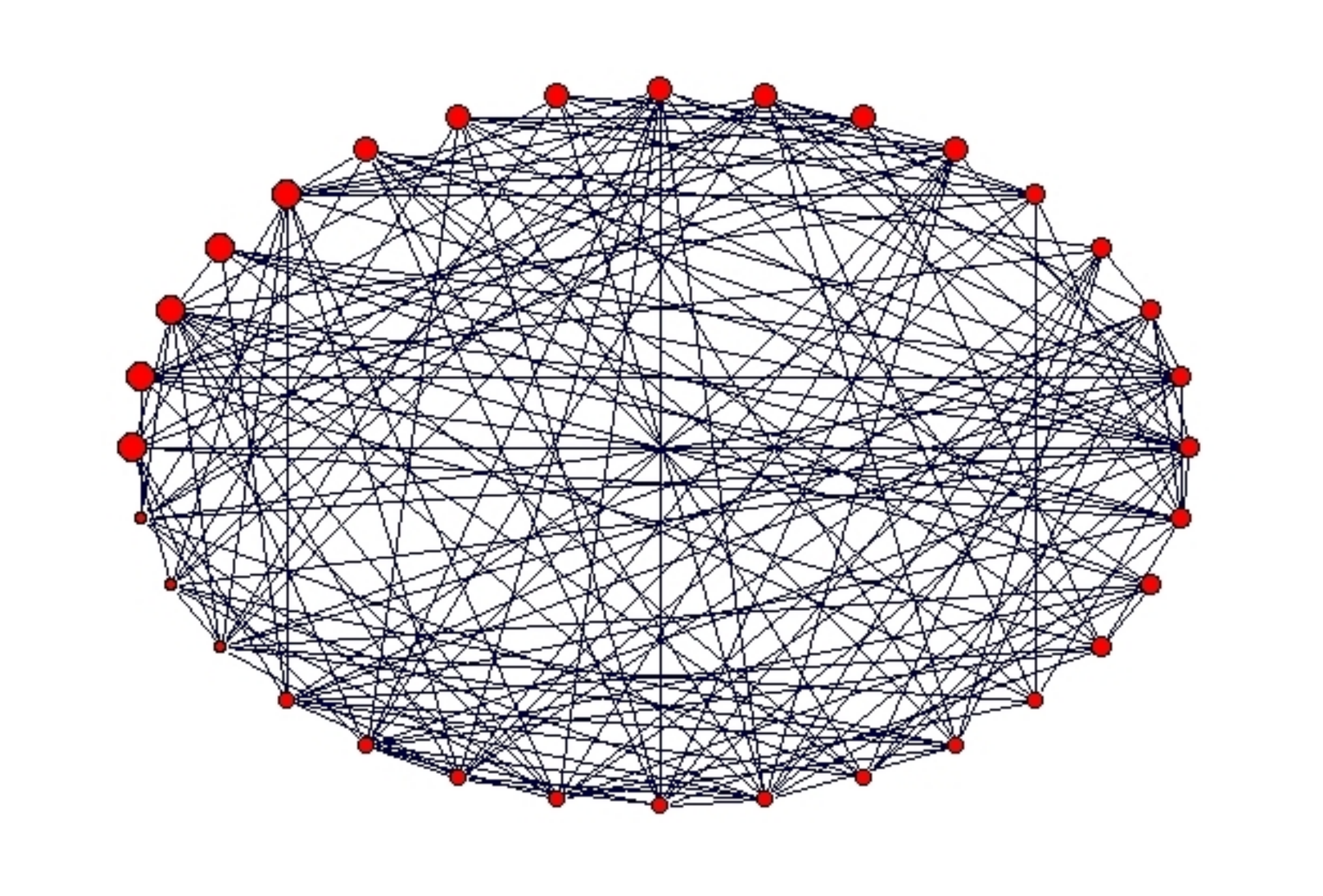}
}
\subfigure[Optimal Network]{
\centering
\includegraphics[width=.45\textwidth]{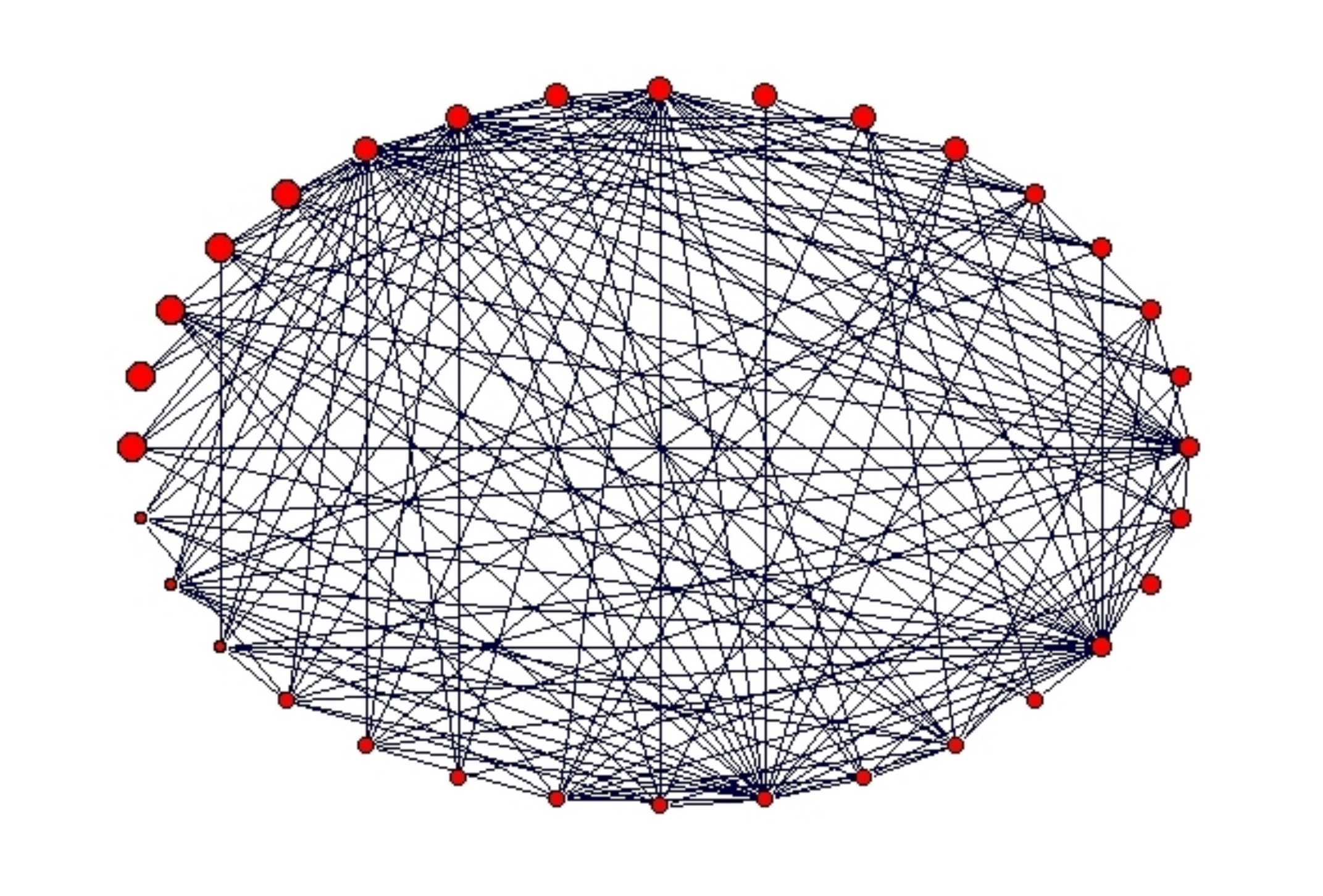}
}
\caption{\label{net_fig}(a) One sample of initial network of 32 vertices and 174 edges is shown. (b) The optimal network obtained from the initial network of (a) is shown. The nearly identical nature is introduced through the NDP $\omega$ and $\omega$ is chose randomly in the interval $(0.9,1.1)$. The node size is proportional to frequency parameter $\omega$, i.e. the node with larger $\omega$ has bigger size.}
\end{figure*}

\subsection{Stable interval in coupling parameter $l_\varepsilon$}

When the variations in the NDP are small the master stability function can be approximated as a linear function near the bifurcation points $\alpha_l$ and $\alpha_h$ and thus one can write
\begin{eqnarray}
\alpha_l' & = & \alpha_l + b_l \nu_l \nonumber \\
\alpha_h' & = & \alpha_h + b_h \nu_h \nonumber
\end{eqnarray}
where, $\nu_l$ and $\nu_h$ are the mismatch parameters corresponding to the eigenvalues $\gamma_2$ and $\gamma_N$ of the coupling matrix $G^T$ and $1/b_l$ and $1/b_h$ are the slopes of master stability function near the points $\alpha_l$ and $\alpha_h$ respectively.

The interval of the coupling parameter $l_{\varepsilon}$ where the synchronization is stable then can be written as
\begin{eqnarray}
l_{\varepsilon} & = & \mid \frac{\alpha_h'}{\gamma_N} - \frac{\alpha_l'}{\gamma_2}\mid 
                  =  l_{\varepsilon}^0 + \mid \frac{b_h\nu_h}{\gamma_N} - \frac{b_l\nu_l}{\gamma_2}\mid
\label{l_eps_eq}
\end{eqnarray}
where, $l_{\varepsilon}^0$ is the stable interval for coupled identical oscillators. We choose $l_{\varepsilon}$ as the order parameter to construct optimized networks with better synchronizability.

Finally, we consider an example of $x$ component coupled nearly identical R\"ossler oscillators. The dynamical equations are give as
\begin{eqnarray*}
\dot{x}^i & = & -\omega_i y^i - z^i + \varepsilon\sum_j^N g_{ij} x^j \\
\dot{y}^i & = & \omega_i x^i + a y^i \\
\dot{z}^i & = & b + z^i(x^i - c)
\end{eqnarray*}
where, the frequency parameter $\omega$ is the NDP and $a,b$ and $c$ are the other R\"ossler parameters. The MSF for nearly identical R\"ossler oscillators is shown in Fig.~\ref{msfrosomega_colormap}. From Fig.~\ref{msfrosomega_colormap}, we can see that the negative region of the MSF increases as the mismatch parameter $\nu_{\omega}$ increases.

\section{Synchronization optimized networks}

In this section we discuss the construction of optimized networks with better synchronizability for coupled nearly identical oscillators. By better synchronizability we mean that the synchronization is stable for the widest possible interval of the coupling parameter $\varepsilon$, i.e. $l_{\varepsilon}$ is maximum for the optimal networks. We construct the optimized networks with two constraints. The number of vertices and the number of edges of the network are fixed and there are no multiple edges and self loops in the networks. By rewiring the network using Metropolis algorithm we obtain the optimal network. Now, we briefly discuss the Metropolis algorithm for construction of optimized networks with better synchronizability.

Let us start with a connected network of $N$ coupled nearly identical oscillators and $E$ edges and the coupling matrix be $G^{old}$. Let the stable interval of the coupling parameter for this initial network be $l_{\varepsilon}^{old}$ where the value of $l_{\varepsilon}$ is determined using Eq.~(\ref{l_eps_eq}). Now we randomly delete one existing edge and create one new edge at an edge vacancy. Thus, we avoid creating multiple edges and self loops. We reject the resultant network if it is disconnected. Otherwise the stable interval $l_{\varepsilon}^{new}$ for the resultant network is determined. We accept the resultant network if $l_{\varepsilon}^{new} - l_{\varepsilon}^{old} > 0$, otherwise we accept the resultant network with a probability $e^{(l_{\varepsilon}^{new} - l_{\varepsilon}^{old})\beta}$, where $\beta = 1/T$ and $T$ is a temperature-like parameter. This rewiring procedure which defines a Monte Carlo step, is repeated several times. We start with a high value of $T$ ($=1$). $T$ is kept fixed for 1000 Monte Carlo steps or 10 accepted ones, whichever occurs first. Then $T$ is reduced by a certain factor ($T_{factor} < 1$) so that stimulated annealing or slow cooling occurs \cite{Binder_book,Binder_book_2,numerical_recipes}. We keep on repeating this process until there are no more changes during five successive temperature steps.

\begin{figure*}
\subfigure[]{
\centering
\includegraphics[width=.4\textwidth]{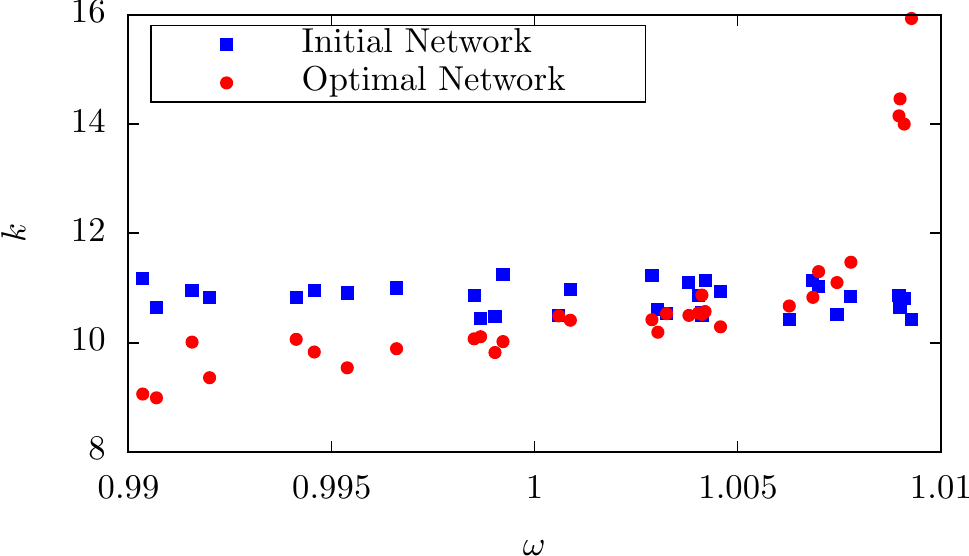}
}
\subfigure[]{
\centering
\includegraphics[width=.4\textwidth]{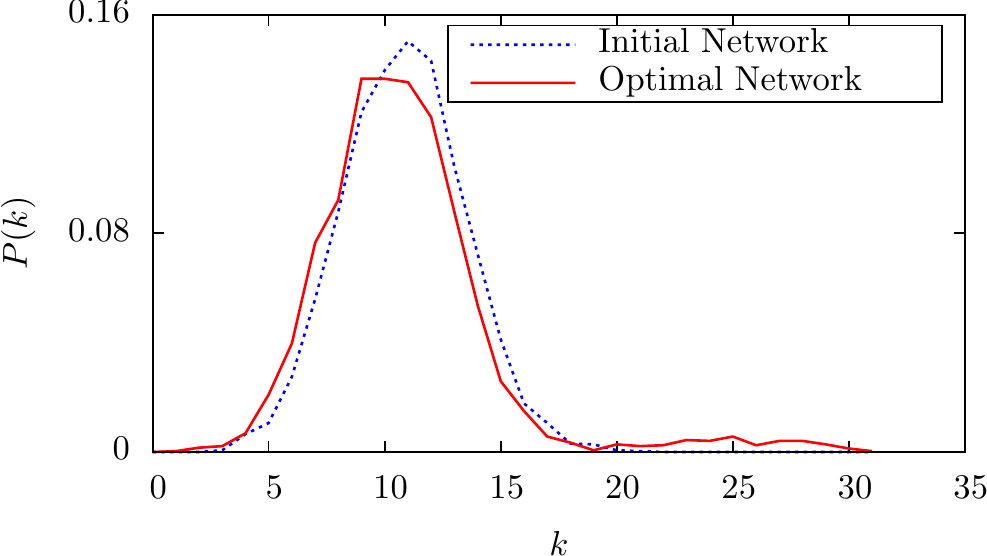}
}
\caption{\label{k_pdist_omega}(a)The degree $k$ of the verices of the initial network (blue squares) and optimized network (red circles) are plotted as a function of the NDP $\omega$ for 32 coupled R\"ossler oscillators. The nodes with higher $\omega$ value have higher degree.
(b)The degree distribution $P(k)$ of the initial network ( blue dotted line) and the optimal network (red solid line) are shown as a function of $k$. For the degree distribution of optimal network a very small peak at higher degree value is seen. These results are averaged over 100 runs.}
\end{figure*}

\section{Structural analysis of optimized networks}

For our numerical simulation we consider an undirected and unweighted network of coupled nearly identical R\"ossler oscillators with $N=32$ and the total number of links $M=174$. The non-identity nature of the coupled R\"ossler oscillators is introduced by considering the frequency parameter $\omega$ as NDP. The frequency parameter $\omega$ is chosen randomly from an interval $(0.9,1.1)$. In Fig.~\ref{leps_rho}(a), the order parameter $l_{\varepsilon}$ is plotted as a function of the Monte Carlo iterations. The stable interval $l_{\varepsilon}$ increases and saturates to a higher value.

In Figs.~\ref{net_fig}(a), a sample of initial network of 32 coupled nearly identical R\"ossler oscillators with 174 edges is shown. The vertex size is proportional to frequency parameter $\omega$, i.e. the vertex with larger $\omega$ has bigger size. In Fig.~\ref{net_fig}(b), the optimal network obtained from the network of Fig.~\ref{net_fig}(a) is shown. In Fig.~\ref{net_fig}(b) we can see that in the optimal network the nodes with higher values of frequencies ($\omega$), have more connections.

In the optimized network we investigate the structural properties of the network. First, we study the vertices which have more connections than other nodes, i.e. the nodes which are selected as hubs. In Fig.~\ref{k_pdist_omega}, we plot the degrees of the vertices as a function of the NDP $\omega$ for the initial network (blue squares) and the optimal network (red circles). For the initial network all vertices have almost similar degree, but in the optimal network the vertices with larger $\omega$ has higher degree than other nodes.

To quantify this effect, we define the Pearson correlation coefficient between the parameter and degree of a node as,
\begin{equation}
\rho_{r k} = \frac{<(k_{i}-<k_i>)(r_{i}-<r_i>)>}{\sqrt{<(k_{i}-<k_i>)^2><(r_{i}-<r_i>)^2>}}
\label{r_k_corr_eq}
\end{equation}
where, $k_i = -g_{ii}$ is the degree of $i$-the node.


\begin{figure*}
\begin{center}
\includegraphics[]{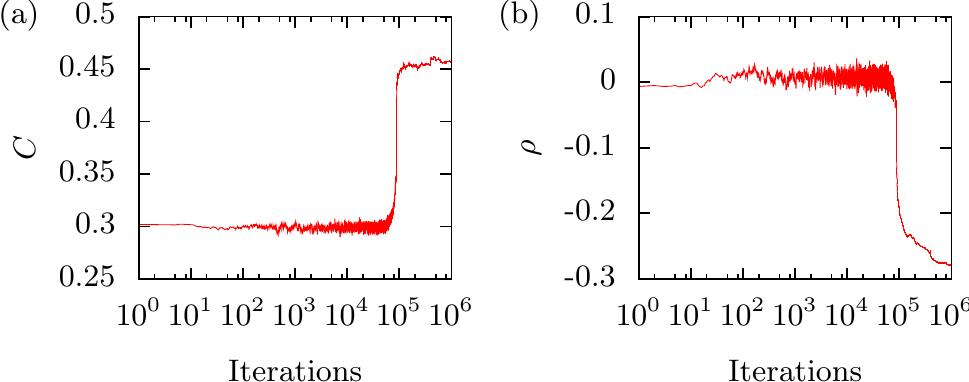}
\end{center}
\caption{\label{cc_r} (a) The average clustering coefficient $C$ is plotted as a function of Monte Carlo iterations. The average clustering coefficient increases for the optimal networks. (b) The degree mixing coefficient $\rho$ is plotted as a function of Monte Carlo iterations.}
\end{figure*}

Fig.~\ref{leps_rho}(b) shows $\rho_{\omega k}$ (solid line) as a function of Monte Carlo steps. For the random network $\rho_{\omega k} = 0$. We find that $\rho_{\omega k}$ increases and saturates to a positive value. Thus, in the synchronized optimized network the nodes which have larger frequencies have more connections and are preferred as hubs. The reason for this is the ``V'' shape of the stability region in Fig.~\ref{msfrosomega_colormap}, i.e. the stability range increases as $\nu_{\omega}$ increases.


The degree distribution $P(k)$ gives the probability that a randomly chosen node will have degree $k$. In Fig.~\ref{k_pdist_omega}(b), the degree distribution $P(k)$ of the initial network (blue dotted line) and the optimal network (red solid line) are shown. The degree distribution of the initial network is Gaussian and has one peak while the degree distribution of the optimal network has a smaller peak at higher degree. The reason for this is the presence of some hubs in the optimal network.

To investigate the question of which edges are preferred, we define the correlation coefficient between the absolute parameter differences between two nodes and the edges as,

\begin{equation}
\rho_{\omega A}= \frac{<(A_{ij} - <A_{ij}>)(|r_i - r_j|-<|r_i - r_j|>)>}{\sqrt{<(A_{ij} - <A_{ij}>)^2><(|r_i - r_j|-<|r_i - r_j|>)^2>}}
\label{r_a_corr_eq}
\end{equation}
where, $A_{ij}=1$ if nodes $i$ and $j$ are connected and 0 otherwise.

Fig.~\ref{leps_rho}(c) shows $\rho_{\omega A}$ as a function of Monte Carlo iterations. We find that $\rho_{\omega A}$ increases from 0 (the value for the random network) and saturates. Thus, in the synchronized optimized network the pair of nodes which have a larger relative frequency mismatch are preferred as edges for the optimized network. Again, the reason for this preference of edges is probably  the conical shape of the stability region in Fig.~\ref{msfrosomega_colormap}. The edges are to be chosen so that the parameter $\nu_{\omega}$ increases and the stability region increases.

The clustering coefficient is another important parameter which quantifies the possibility that two neighbors of a common node are also neighbors. The clustering coefficient $c_i$ of vertex $i$ is defined as
\begin{equation}
c_i = \frac{2 e_i}{k_i(k_i-1)}
\end{equation}
where, $e_i$ is the number of edges that exist among the neighbors of vertex $i$ and $k_i$ is the degree of vertex $i$.  The clustering coefficient $C$ of the entire network is defined as
\begin{equation}
C = \frac{1}{N}\sum_i c_i.
\end{equation}

In Fig.~\ref{cc_r}(a) the clustering coefficient $C$ of the network is plotted as a function of the Monte Carlo iterations. From Fig.~\ref{cc_r} we can see that the clustering coefficient of the network increases and saturates to a higher positive value. Thus, the optimized network has more local structure than the random network, i.e. there are more triangles than the random network. The result is intuitively easy to understand. Forming a loop will enhance the stability of synchronization due to a faster feedback and smaller the size of the loop better will be the result. We note that the behavior is similar to that for coupled identical oscillators where it has been noticed that networks with larger value of clustering coefficient have better stability of synchronization \cite{Donetti2005}.

Assortative mixing in networks~\cite{NewmanPRL2002,NewmanPRE2003} gives the tendency of vertices to be connected with vertices of comparable degrees. Let the degrees of verices at the ends of the $i$th edge connecting vertices $j$ and $l$ be $(k_j)_i$ and $(k_l)_i$. Following Ref~\cite{NewmanPRL2002} the degree mixing coefficient $\rho$ can be calculated as
\begin{equation}
\rho = \frac{\frac{1}{M} \sum_i (k_j)_i (k_l)_i - \left[\frac{1}{2M} \sum_i ((k_j)_i+(k_l)_i)\right]^2}{\frac{1}{2M}\sum_i ((k_j)_i^2 + (k_l)_i^2) - \left[\frac{1}{2M} \sum_i ((k_j)_i+(k_l)_i)\right]^2}, \label{assortative_coefficient}
\end{equation}
where, $M$ is the total number of edges in the network and the sums are over the edges. When comparable degree nodes get connected the correlation coefficient $\rho$ is positive and the network is called assortative network. The network is called disassortative network when the coefficient $r$ is negative. This happen when high degree vertices get connected with low degree vertices. For networks which show no assortative mixing the correlation coefficient $\rho$ is zero. The random networks of Erd\H{o}s and R\'{e}nyi and the scale free network model of Barab\'{a}si and Albert shows no assortative mixing. It has been observed that many naturally evolving networks, such as internet, WWW, protein interaction, neural networks, etc. shows disarrortative mixing of degree~\cite{NewmanPRL2002,ChavezPRE2006,Bernardo2005,Sorrentino2007}.

In Fig.~\ref{cc_r}(b) the degree mixing correlation coefficient $\rho$ is shown as a function of Monte Carlo iterations. The degree mixing coefficient $\rho$ decreases and becomes negative. Thus the optimized network is disassortative in nature. This behavior is consistent with that of coupled identical oscillators..

\section{Conclusion}

To conclude we have extended the MSF formalism to analyze stability of generalized synchronization for nearly identical oscillators. Using the stability criteria given by the MSF we construct optimal networks with better synchronizability by using Metropolis algorithm. We find the hubs of the optimal networks are nodes with larger frequency. The optimal network is disassortative in nature, i.e. nodes with higher degree tend to connect with node with lower degree.

\end{document}